%% file: ms.tex
\documentclass[10pt]{article}
\usepackage{verbatim,color,amssymb,epsfig}
\usepackage{amsmath,amsthm,graphicx,bbm,url}
\usepackage[usenames,dvipsnames]{xcolor}
\usepackage{fancyhdr}
\usepackage[authoryear, sort]{natbib}
\usepackage{subfig}
\usepackage{enumitem}
\usepackage{float}

\usepackage{mathtools}
\DeclarePairedDelimiter\ceil{\lceil}{\rceil}
\DeclarePairedDelimiter\floor{\lfloor}{\rfloor}

\newcommand{\INFTY}{\infty}

\newcommand{\E}{\mathbb{E}}

\newcommand{\ind}[1]{\mathbbm{1}_{#1}}

\usepackage{xspace}

\newcommand{\Revision}[1]{#1}

\newtheorem{Thm}{\underline{\bf Theorem}}

\def\E{\mathbb{E}}

\begin{document}
\thispagestyle{empty}
\baselineskip=28pt
\begin{center}
{\LARGE{\bf A Study of Functional Depths}}
\end{center}

\baselineskip=12pt

\vskip 2mm
\begin{center}
James P. Long\\
Department of Statistics, Texas A\&M University\\
3143 TAMU, College Station, TX 77843-3143\\
jlong@stat.tamu.edu\\
\hskip 5mm \\
Jianhua Z. Huang\\
Department of Statistics, Texas A\&M University\\
3143 TAMU, College Station, TX 77843-3143\\
jianhua@stat.tamu.edu\\
\end{center}

\hskip 5mm

\begin{center}
{\Large{\bf Abstract}}
\end{center}
\baselineskip=12pt

Functional depth is used for ranking functional observations from most outlying to most typical. The ranks produced by functional depth have been proposed as the basis for functional classifiers, rank tests, and data visualization procedures. Many of the proposed functional depths are invariant to domain permutation, an unusual property for a functional data analysis procedure. Essentially these depths treat functional data as if it were multivariate data. In this work, we compare the performance of several existing functional depths to a simple adaptation of an existing multivariate depth notion, $L^\infty$ depth ($L^{\infty}D$). On simulated and real data, we show $L^{\infty}D$ has performance comparable or superior to several existing notions of functional depth. In addition, we review how depth functions are evaluated and propose some improvements. In particular, we show that empirical depth function asymptotics can be mis--leading and instead propose a new method, the rank--rank plot, for evaluating empirical depth rank stability.

\baselineskip=12pt
\par\vfill\noindent

\par\medskip\noindent
\underline{\bf Short title}: A Study of Functional Depths

\clearpage\pagebreak\newpage
\pagenumbering{arabic}
\newlength{\gnat}
\setlength{\gnat}{22pt}
\baselineskip=\gnat

\section{Introduction} 
\label{sec:intro}

Given a set of observations $x_1, \ldots, x_n$ in $\mathbb{R}^p$, a depth function $\mathcal{D}:\mathbb{R}^p \rightarrow \mathbb{R}^1$ provides a center--outward ordering for the data. Observations with large depth are near the ``center'', low depth observations are outliers. The ordering induced by $\mathcal{D}$ has been used to identify outliers and medians, perform robust mean estimation in the presence of outliers, and construct statistical classifiers \citep{cuevas-fraiman-dual,li2012dd,donoho-gasko,liu-parel-sin}. 

With the increased importance of functional data, various works have developed new depth functions for the infinite dimensional setting. Let $\mathcal{F}$ be some function space. A functional depth $\mathcal{D}$ maps $\mathcal{F}$ to $\mathbb{R}^1$. A primary motivation for these new depths has been computational tractability; many depths for multivariate data were not computationally feasible in dimension greater than 10 (for example Tukey depth \citep{tukey-depth} and Simplicial depth \citep{liu}). Recent proposals for functional depths include the band depth (BD) and modified band depth (MBD) \citep{lopez2009concept}, half region depth (HRD) and modified half region depth (MHRD) \citep{lp-r-half}, random Tukey depth (RTD) \citep{cuesta2008random}, and spatial depth (SPATD) \citep{chakraborty2014data,sguera2014spatial}. BD, MBD, HRD, and MHRD were designed specifically for functional data while RTD and SPATD are extensions of multivariate depths procedures to the functional setting. As with multivariate depths, there are many proposed uses for functional depth: exploratory data analysis (eg functional boxplots \citep{genton-sun-functional-box}), inference (eg rank tests \citep{lopez2009concept}), and prediction (eg classifiers \citep{cuevas2007robust}).

\Revision{Functional data is often stored in an $n \times p$ matrix where $n$ is the number of functions and $p$ is the number of time points at which the functions are sampled. All the above functional depths are invariant to domain permutation: reordering the $p$ columns of the data matrix will not change the resulting depths. This contrasts with most functional data analysis procedures, such as fitting splines to each function. Invariance to domain permutation is not an inherently bad property.\footnote{See \cite{lopez2007functional} for further discussion of domain permutation invariance.}} However this observation suggests that computationally tractable multivariate depths may be competitive with data depths developed specifically for functional data, such as BD, MBD, HRD, and MHRD. In this work we compare the performance of several functional depths to $L^{\infty}$ depth ($L^{\infty}D$), a multivariate depth applied to functional data. With $L^{\infty}D$, the depth at $x$ is inversely proportional to the mean $L^{\infty}$ distance to all observations. $L^{\infty}D$ was proposed, but never extensively studied, in a work by \cite{serfling-zuo-notions}. As a means of comparing depths we follow a standard set of criteria for evaluating depth performance. These include:
\begin{enumerate}
\item \underline{Depth Function Properties}: \cite{serfling-zuo-notions} proposed four properties that a multivariate depth function should satisfy. Proposals for functional depth notions have checked to what extent the proposed depth possesses these properties. Additional properties, such as non--degeneracy of the depth function, have been studied by \cite{chakraborty2014data} and \cite{kuelbs2014half}.\label{num:properties}
\item \underline{Convergence of Estimator}: Supposing a particular depth function is useful for some application, it still must be estimated from the data. Various works have proven law of large numbers and central limit theorems for depth function estimators. \label{num:conv}
\item \underline{Computation}: Many multivariate depth functions are computationally intractable in dimension greater than 10. Since functional data is generally stored in a computer as a dense grid of points, any functional depth must be computable on the size of the grid. Additionally for moderate and large data sets, computation of depth must scale well in the number of observations.\label{num:comp}
\item \underline{Applications}: Functional depth has many potential applications. Depths functions may be compared by how well they perform these tasks. \label{num:app}
\end{enumerate}

We find that $L^{\infty}D$ has performance comparable to existing notions of functional depth as measured by the standards of 1--3 above. While it is infeasible to test $L^{\infty}D$ on every possible application, we find it has good performance in a standard set of robust mean estimator simulations. In a real data application from astronomy $L^{\infty}D$ identifies outlying functions which have been mis--registered. On the whole, we find that $L^{\infty}D$ has performance superior to BD, HRD, MHRD, and RTD and comparable to MBD and SPATD.

A second purpose of this work is to critique some of the above criteria and propose some improvements.  In particular, with regard to convergence of estimators (criteria \ref{num:conv}), little attention has been paid to when asymptotics take hold or how results such as the Law of Large Numbers for an empirical depth function apply to the depth rankings, the quantity of primary interest in many applications. In Section \ref{sec:depth_rankings_conv} we show that asymptotics can give a misleading picture about the depth ranks. We propose a new method, the rank--rank plot, for evaluating the convergence and stability of depth rankings. On a real data set, the rank--rank plot reveals interesting phenomenon for HRD, MHRD, and RTD.

As part of this work, we make publicly available a set of robust mean estimation simulations which may be used to compare different notions of functional depth. These models, and similar versions, have been used in the past by \cite{lopez2009concept}, \cite{lp-r-half}, and \cite{fraiman2001trimmed}. Additionally we have curated a data set of periodic variable stars collected from astronomy databases. This functional data set may be used to test new and existing functional depths.

This paper is organized as follows. In Section \ref{sec:linf} we introduce $L^{\infty}D$ and discuss its performance on criteria \ref{num:properties}, \ref{num:conv}, and \ref{num:comp} in comparison to existing notions of functional depth. In Sections \ref{sec:sim} and \ref{sec:lc} we compare the performance of $L^{\infty}D$ to other depths on problems involving robust mean estimation and identification of outliers and medians in an astronomy data set. These sections give an indication of how $L^{\infty}$ performs in applications (criteria \ref{num:app}). We conclude in Section \ref{sec:conclude}. Code and data for reproducing all results are described in Section \ref{sec:supp} and available online.

\section{$L^{\infty}$ Depth}
\label{sec:linf}

\cite{serfling-zuo-notions} defined multivariate $L^p$ depth functions to be of the form $L^pD(x,P) = (1 + \mathbb{E}[||x-X||_{p}])^{-1}$ where $x \in \mathbb{R}^p$ and $P$ is some distribution on $\mathbb{R}^p$. \cite{serfling-zuo-notions} verified that $L^{p}$ depth satisfied certain properties in the multivariate case. They did not study its applications on data, either multivariate or functional. They also did not compare its properties to any functional depths.

We now provide a straightforward generalization of $L^{\infty}D$ to the functional case. Let $I$ be some compact interval of $\mathbb{R}$ and $C(I)$ be the set of continuous functions on $I$. Let $X = \{X(t) : t \in I\}$ be a process in $C(I)$ with distribution $P$. For any functions $x,y \in C(I)$ define $||x-y||_{\infty} = \sup_{t \in C(I)}|x(t) - y(t)|$. The $L^{\infty}$ depth for functional data is
\begin{equation}
\label{eq:linfp}
L^{\infty}D(x,P) = (1 + \mathbb{E}[||x-X||_{\infty}])^{-1}.
\end{equation}
For an independent sample $X_1, \ldots, X_n$ from $P$, the empirical $L^{\infty}D$ is
\begin{equation}
\label{eq:linfpn}
L^{\infty}D(x,P_n)= (1 + n^{-1}\sum_{i=1}^n||x-X_i||_{\infty}])^{-1}.
\end{equation}
A function $x$ has low depth if $n^{-1}\sum_{i=1}^n||x-X_i||_{\infty}$ is large, signalling it is far away on average from the sample functions. A function $x$ has large depth if $n^{-1}\sum_{i=1}^n||x-X_i||_{\infty}$ is small, signalling it is close on average to the sample functions.

As a direct generalization of a multivariate depth to functional data, $L^{\infty}D$ does not use any properties inherent to functional data such as smoothness. \Revision{In this work we compare $L^{\infty}D$ to six other functional depths: band--depth with $J=3$ functions delimiting bands (BD), modified band--depth with $J=2$ functions delimiting bands (MBD), half--region depth (HRD), modified half--region depth (MHRD), random Tukey depth with $250$ random projections (RTD), and spatial depth (SPATD). We use $J=3$ delimiting bands for BD because this is recommended in the work of \cite{lopez2009concept}. We use $J=2$ delimiting bands for MBD because this is the most popular form \citep{kwon2016clustering,genton-sun-functional-box,sun2012exact}. We use $250$ random projections for RTD because it has been suggested that this is sufficient in a wide range of cases (see \cite{cuesta2008random}).} In the remainder of this section, we evaluate functional $L^{\infty}D$ with respect to criteria \ref{num:properties}, \ref{num:conv}, and \ref{num:comp} above. We discuss $L^{\infty}D$ performance in applications in Sections \ref{sec:sim} and \ref{sec:lc}.

\subsection{Depth Function Properties}

Following Tukey's half space depth, many other depths were proposed for multivariate data. In an attempt to standardize the definition of depth, \cite{serfling-zuo-notions} proposed four properties that a depth function must satisfy. Roughly these properties are affine invariance, maximality at center, monotonicity relative to deepest point, and vanishing at infinity. There is disagreement about how important these properties are. For example, \cite{baggerly1999discussion} critique the maximality at center property in the context of multimodal data.

Recently these four properties have been used to motivate and justify new depths for functional data. As in the multivariate case, it is not universally accepted that these properties are important for a depth function. Recent work has proposed new properties for functional depths \citep{nieto2014definition}.

\cite{serfling-zuo-notions} (Corollaries 2.3 and 2.4) showed that $L^{\infty}D$ applied to multivariate data satisfies three of the four properties (fails affine invariance). Here we show that in the functional case $L^{\infty}D$ still satisfies these properties. Let $D(\cdot,\cdot)$ be some depth function. The four properties of \cite{serfling-zuo-notions}, adapted to functional data, are:

\begin{itemize}[label={}]
\item \Revision{P1) Affine Invariance: Let $A:\mathcal{F} \rightarrow \mathcal{F}$ be an invertible, bounded linear operator. Then for all $x,y \in \mathcal{F}$
\begin{equation*}
D(x,P) = D(A(x) + y,P_{A(X)+y})
\end{equation*}
where $P_{A(X)+y}$ is the probability measure of the process $A(X)+y$.}
\item P2) Maximality at Center: If $P$ is a distribution with center $\theta$ (i.e. $X-\theta = \theta - X$ in distribution), $D(\theta,P) = \sup_{x} D(x,P)$.\footnote{We use \cite{serfling-zuo-notions} notion of C--symmetry for defining center.}
\item P3) Monotonicity Relative to Deepest Point: For any $P$ having deepest point $\theta$,
\begin{equation*}
D(x,P) \leq D(\theta + \alpha(x-\theta),P)
\end{equation*}
holds for all $\alpha \in [0,1]$.
\item P4) Vanishing at Infinity: $D(x,P) \rightarrow 0$ as $||x||_{\infty} \rightarrow \infty$.
\end{itemize}

\begin{Thm}
\label{thm:properties}
If $P$ is such that $\mathbb{E}[||X||_{\infty}]$ exists, then $L^{\infty}D$ satisfies P2, P3, and P4. \Revision{In addition $L^{\infty}D$ is invariant up to isometric transformations, ie if $||A(x)||_{\infty} = ||x||_\infty \, \, \forall x$ then $L^{\infty}(x,P) = L^\infty(A(x),P_{A(X)})$.}
\end{Thm}

A proof of Theorem \ref{thm:properties} is given in Section \ref{prf:properties}. \cite{lopez2009concept} (Theorem 1) showed that multivariate band depth satisfies P2, P3, and P4. Functional BD is shown to satisfy P4 (Theorem 3, Part 3).  \cite{lp-r-half} showed half region depth satisfies P4. RTD satisfies properties 2 and 3 (Theorem 2.9 of \cite{cuesta2008random}). Thus we find that $L^{\infty}D$ is competitive relative to other functional depths when evaluated based on how well it satisfies these four properties.

\subsection{Convergence of Estimator}
\label{sec:conv}

The properties of \cite{serfling-zuo-notions} concern population level depth functions. In practice, empirical depths are used to approximate the population depth function. It is straightforward to show a pointwise strong law and pointwise Central Limit Theorem for $L^{\infty}D(x,P_n)$. 

\begin{Thm}[Strong Law of Large Numbers]
\label{thm:slln}
Let $P$ and $x \in C(I)$ be such that $\E[||X - x||_{\infty}] < \infty$. Then
\begin{equation}
  L^{\infty}D(x,P_n) \rightarrow_{a.s.} L^{\infty}D(x,P).
\end{equation}
\end{Thm}

\begin{Thm}[Central Limit Theorem]
Let $P$ and $x \in C(I)$ be such that $\E[||X - x||_{\infty}] = \mu_x < \infty$ and $Var(||X - x||_{\infty}) = \sigma^2_X < \infty$. Then
\label{thm:clt}
\begin{equation}
\label{eq:clt}
\sqrt{n}(L^{\infty}D(x,P_n) - L^{\infty}D(x,P)) \rightarrow_{d} N(0,\sigma_x^2(1+\mu_x)^{-4}).
\end{equation}
\end{Thm}

Proofs of these theorems are given in Appendix Sections \ref{prf:slln} and \ref{prf:clt}. Convergence results have been proven for other functional depths. For example \cite{lopez2009concept} proved a uniform strong law for band depth in the multivariate case and a uniform strong law across any equicontinuous set in the functional case. For half region depth, \cite{lp-r-half} showed a uniform strong law across sets with finite bracketing number. \cite{cuesta2008random} proved a uniform strong law of large numbers for random Tukey depth (Theorem 2.10). 

While these results provide some assurance of the quality of the empirical depth approximation, they do not indicate how well the empirical depth ordering of the observations approximates the population depth ordering of the observations. Since in most depth applications all that is used is the ordering of the observations, understanding this approximation may be more important than convergence results for the depth function. Further, the asymptotic nature of these results does not directly address what happens at finite sample sizes. In Section \ref{sec:depth_rankings_conv} we propose the rank--rank plot for addressing these issues.

\subsection{Computation}

In most depth applications, the depths of all observations are computed using the empirical depth function. A major motivation for the creation of functional depths is that common multivariate depths, such as Tukey and Simplicial, could not perform this task in a reasonable amount of time in high dimensions. $L^{\infty}D$ (and more generally $L^p$ depth) is $O(n^2d)$ for computing the empirical depths of all observations where $n$ is the number of sample functions and $d$ is the number of points at which the functions have been sampled. Using \texttt{R} implementations of all depths, we find that $L^{\infty}D$ is faster than all depths except MBD. MBD is extremely fast due to an algorithm of \cite{sun2012exact}. BD and RTD are the slowest. We refer to Appendix \ref{sec:computational_speed} for a more detailed discussion on computational speed.

Having established some properties of $L^{\infty}D$ function, we now explore its performance on a set of robust mean estimation simulations (Section \ref{sec:sim}) and an astronomy data set (Section \ref{sec:lc}).


\section{Application: Robust Mean Estimation and Outlier Identification}
\label{sec:sim}

A common application of depth is robust mean estimation in the presence of outliers. Given a set of observations, one can order them using the empirical depth function and compute a mean using the $(1-\alpha)$ deepest functions where $0 \leq \alpha < 1$. This procedure is a generalization of trimmed means to the functional data setting. Depths may be compared based on how well they estimate the mean. Here we compare the performance of $L^{\infty}D$ to other depth measures using a standard set of magnitude and shape outlier models.

\subsection{Magnitude Contamination}

\begin{figure}[t]
  \begin{center}
    \begin{includegraphics}[height=3in,width=6in]{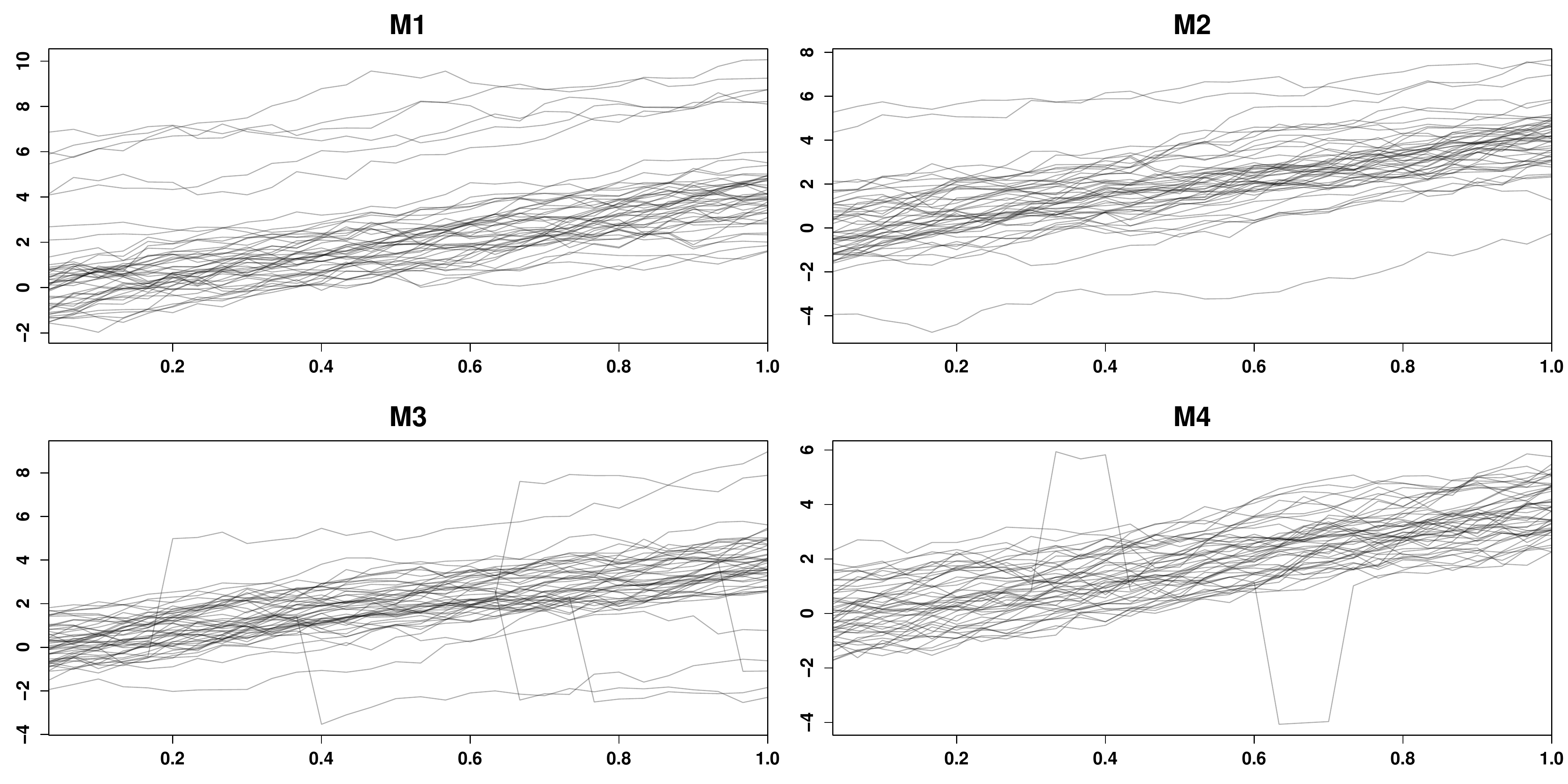}
      \caption{Magnitude outlier models with $M=5$.\label{fig:plot_models_mag}}
    \end{includegraphics}
  \end{center}
\end{figure}

We use five magnitude outlier models, M0--M4, used in \cite{lp-r-half}. Similar models were used in \cite{lopez2009concept} and \cite{fraiman2001trimmed}. For each model the true underlying function to be estimated is $f(t) = 4t$ for $t \in [0,1]$ and the samples are of size $n=50$. In M0, the no outlier model, observations are $X_i(t) = f(t) + e_i(t)$  where $e_i$ is a mean $0$ Gaussian process with covariance $\E[e_i(t)e_i(s)] = \exp\{-|t-s|\}$. In models M1 through M4 the base model M0 is contaminated with outliers. Let $\epsilon_i \sim Bern(q)$, $\sigma_i$ be $-1$ with probability $1/2$ and $1$ with probability $1/2$, $T_i \sim Unif(0,1)$, and $M > 0$ be some constant. The models are:
\begin{itemize}[label={}]
\item M0) No contamination: 
\begin{equation*}
Y_i(t) = X_i(t)
\end{equation*}
\item M1) Asymmetric total contamination: 
\begin{equation*}
Y_i(t) = X_i(t) + \epsilon_iM.
\end{equation*}
\item M2) Symmetric total contamination: 
\begin{equation*}
Y_i(t) = X_i(t) + \epsilon_i\sigma_iM.
\end{equation*}
\item M3) Partial contamination: 
 \begin{displaymath}
   Y_i(t) = \left\{
     \begin{array}{lr}
       X_i(t) + \epsilon_i\sigma_iM, &  t \geq T_i\\
       X_i(t), &  t < T_i
     \end{array}
   \right.
\end{displaymath} 

\item M4) Peak contamination: 
 \begin{displaymath}
   Y_i(t) = \left\{
     \begin{array}{lr}
       X_i(t) + \epsilon_i\sigma_iM, & t \in [T_i,T_i + l]\\
       X_i(t), & t \notin [T_i,T_i + l]
     \end{array}
   \right.
\end{displaymath} 
where $l = 2/30$.
\end{itemize}
Figure \ref{fig:plot_models_mag} shows a set of $n=50$ observations from models M1-M4 with $M=5$ and $q=0.1$. We assess performance of an estimate $\widehat{f}$ using integrated squared error (ISE) approximated at $L=30$ points in $[0,1]$. For $\widehat{f}$ the ISE is
\begin{equation*}
ISE = \sum_{k=1}^L (\widehat{f}(k/L) - f(k/L))^2.
\end{equation*}
For each estimator and model, we report an estimate of the mean integrated squared error (MISE), computed by averaging the $ISE$ over $N=500$ runs as well as the standard error of the MISE estimate. We use nine estimators. MEAN and MEDIAN are computed by taking the pointwise mean and median of each sample. For the depth based estimators, the functions are ranked using depth and the $\alpha=0.2$ least deep functions are trimmed before computing a mean on the $1-\alpha$ remaining fraction of curves. In other words, using the empirical depth $\mathcal{D}(\cdot,P_n)$, observations are ranked $r_1,\ldots, r_n$ (least deep to deepest) where 
\begin{equation*}
r_i = \sum_{j=1}^n \ind{\mathcal{D}(Y_j,P_n) \leq \mathcal{D}(Y_i,P_n)}.
\end{equation*}
The $\alpha$ trimmed robust mean estimate is
\begin{equation*}
\bar{Y}_{\alpha} = \frac{\sum_{i=1}^n Y_i \ind{r_i > n\alpha}}{\ceil{n(1-\alpha)}}.
\end{equation*}

\input{figs/mag_M5_table}

\input{figs/mag_M25_table}

Results for $M=5$ are contained in Table \ref{figs/mag_M5} and results for $M=25$ are contained in Table \ref{figs/mag_M25}. The two best performing methods for each model are highlighted in bold. $L^{\infty}D$ performs well relative to other depth functions. For model $M0$ the mean is the best estimator with both $M=5$ and $M=25$ level of contamination. \Revision{$L^{\infty}D$ depth performs especially well for the asymmetric contamination model $M1$ because the functions shifted by $M$ are consistently identified as outliers. Other methods, such as HRD and RTD, identify some of the functions shifted by $M$ and some of the functions just below $f(t) = 4t$ as outliers, resulting in a large MISE.} $L^{\infty}D$ provides more improvement over other depths for the $M=25$ models than the $M=5$ models. SPATD and $L^{\infty}D$ are the most effective depths for these simulations.

\subsection{Shape Contamination}
\label{sec:shape}

\begin{figure}[t]
  \begin{center}
    \begin{includegraphics}[height=4.5in,width=6in]{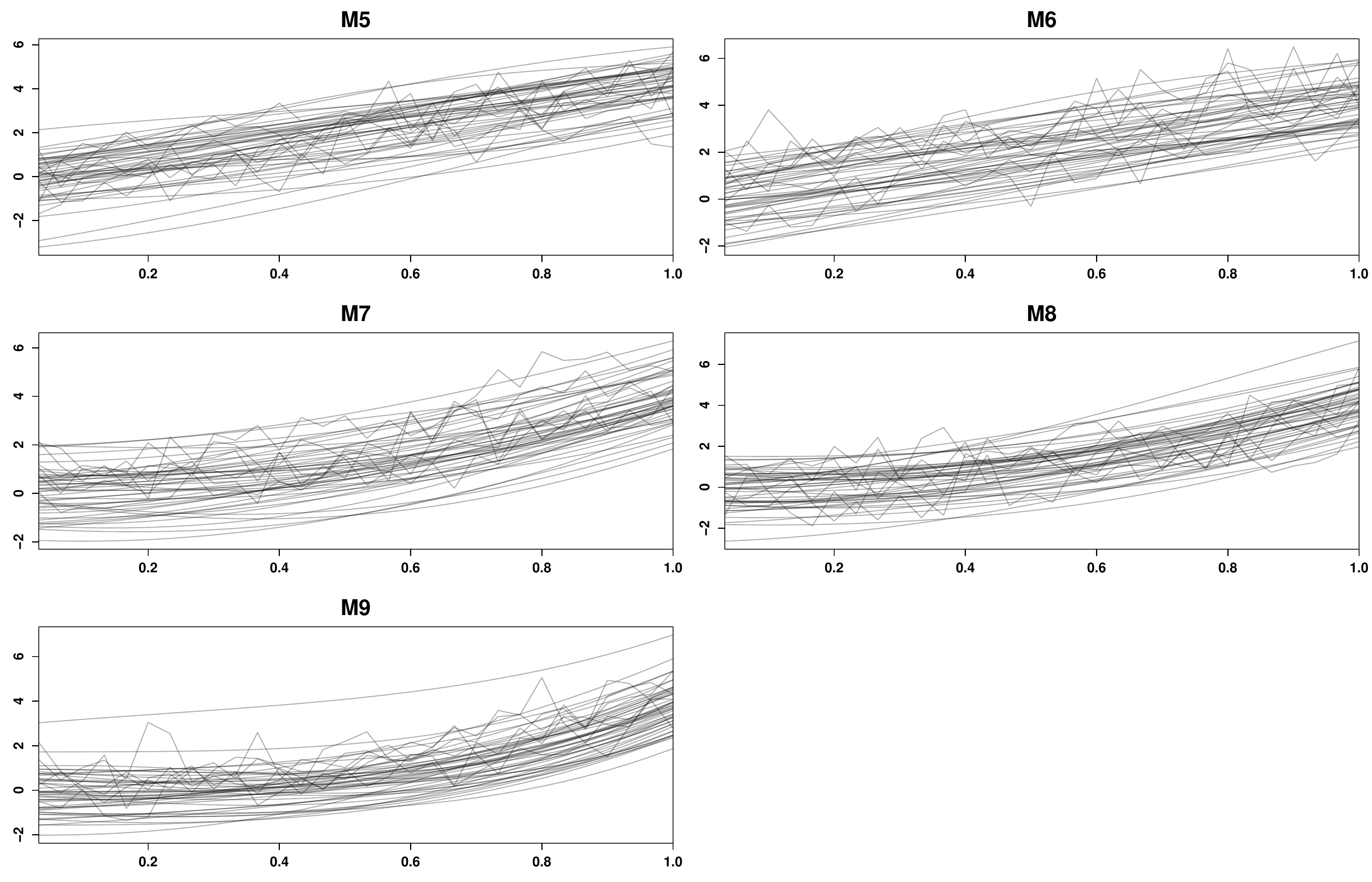}
      \caption{Shape outlier models.\label{fig:plot_models_shape}}
    \end{includegraphics}
  \end{center}
\end{figure}

Following \cite{lp-r-half} we also study $5$ shape outlier models. Here the outliers have different shapes than the non--outliers but the same means. A set of $50$ realizations from each model is shown in Figure \ref{fig:plot_models_shape}. One can see that the outliers will not have a large impact on mean estimation because the outliers are no further from the process mean than many of the non--outliers. Thus we may suspect that a depth function which successfully identifies outliers may not improve mean estimation by a large amount.

First we describe the models. The base model is $X_i(t) = f(t) + e_{1i}(t)$ while the outliers are $Y_i(t) = f(t) + e_{2i}(t)$. For models M5 and M6, $f(t) = 4t$. For models M7 and M8, $f(t) = 4t^2$. For model M9, $f(t) = 4t^3$. The error term $e_{1i}$ is a mean $0$ Gaussian process with covariance $\gamma_1(s,t) = \exp{(-|t-s|^2)}$. The error term $e_{2i}$ is a mean $0$ Gaussian process with covariance $\gamma_2(s,t) = \exp{(-|t-s|^{\mu_2})}$ where $\mu_2 < 2$ so that $e_{21}$ is less smooth than $e_{1i}$. For models M5 and M7, $\mu_2=0.2$ while for models M6, M8, and M9, $\mu_2 = 0.1$. For each model we observe $Z_i(t) = (1-\epsilon_i)X_i(t) + \epsilon_iY_i(t)$ where $\epsilon_i \sim Bern(q)$ with $q=0.15$. These models are meant to exactly match models M5 -- M9 in \cite{lp-r-half}.

\input{figs/shape_table}

As with the magnitude contamination models, we assess performance for the shape models using MISE. Table \ref{figs/shape} contains MISE and standard errors for the 5 models and 9 estimators. Here we see that the mean is always the best and modified band depth is usually the second best. $L^{\infty}D$ performs poorly for these models.

\input{figs/detect}

The fact that MBD is the best depth for mean estimation does not imply that MBD is successfully identifying outliers. When examining the shape models in Figure \ref{fig:plot_models_shape} one can see that non--outliers (smooth functions) that are far from the mean (either shifted above or below the mean) will have more effect on the mean estimate than the less smooth outliers which are sometimes above and sometimes below the mean. To test which methods are identifying outliers most often, we generate 49 curves from the base model of models M5 -- M9 and 1 curve from the outlier model of models M5 -- M9. We repeat this process 500 times and compute the fraction of times that each depth function identifies the outlier to be among the least $0.2$ least deep curves. Results are contained in Table \ref{fig:detect}.

We see that $L^{\infty}D$ followed by BD are the best methods for identifying the outlier. These depths did not perform well on robust mean estimation with the shape models. In contrast, MBD, which performed well for mean estimation, performs poorly at identifying the outliers. This shows that the best depth function depends on the inference goal (mean estimation versus outlier detection), not just the underlying distribution.

\Revision{In these models, the outliers are far from each of the smooth curves at some time point, although not on average. $L^{\infty}D$ works well at identifying outliers here because it is sensitive to maximum, not average, separation between functions. One can consider the opposite setting where most functions are rough and the outliers are smooth. Our real data example in Section \ref{sec:lc} is an example of this situation. In this setting SPATD appears better at detecting outliers than $L^{\infty}D$.}

\section{Application: Outliers and Medians in Astronomical Light Curves}
\label{sec:lc}

Periodic variables are stars that vary in brightness periodically over time. For a given periodic variable star, astronomers measure the brightness many times. Let $t_k$ be the time of brightness measurement $m_k$. Astronomers collect $\{(t_k,m_k)\}_{k=1}^K$ where $K$ is the number of brightness measurements for a particular star. Using period estimation methods, astronomers determine a period for the star. The pattern of brightness variation is evident when one views the brightness as a function of phase ($= $ time modulo period), known as the folded light curve.

\begin{figure}[ht]
  \begin{center}
    \begin{includegraphics}[height=2in,width=6in]{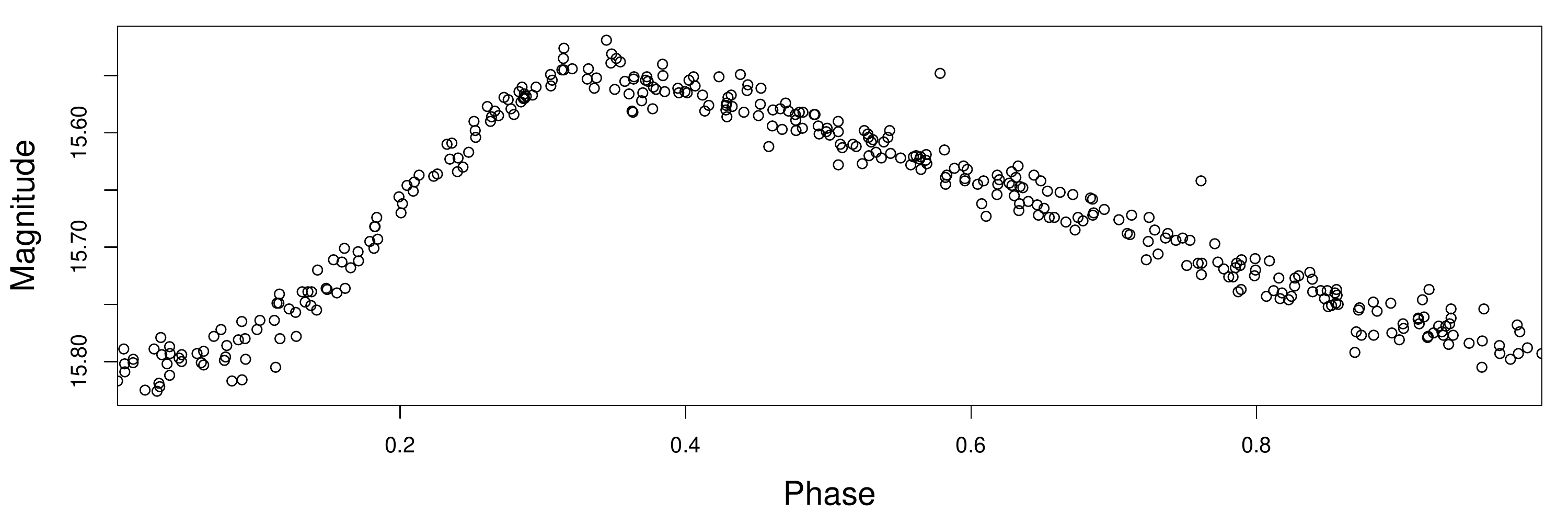}
      \caption{Folded light curve of a Classical Cepheid star from the OGLE-III survey.\label{fig:lc}}
    \end{includegraphics}
  \end{center}
\end{figure}

Figure \ref{fig:lc} shows the folded light curve of a periodic variable of the type Classical Cepheid. Magnitude is inversely proportional to brightness, so larger magnitudes are plotted lower on the y--axis. We study differences in lightcurve shape so we normalize the mean magnitude of each star to $0$. Additionally the phase of the light curve is arbitrary. Thus when comparing shapes of many stars it is essential to align phases.
\begin{figure}[ht]
  \begin{center}
    \begin{includegraphics}[height=2in,width=6in]{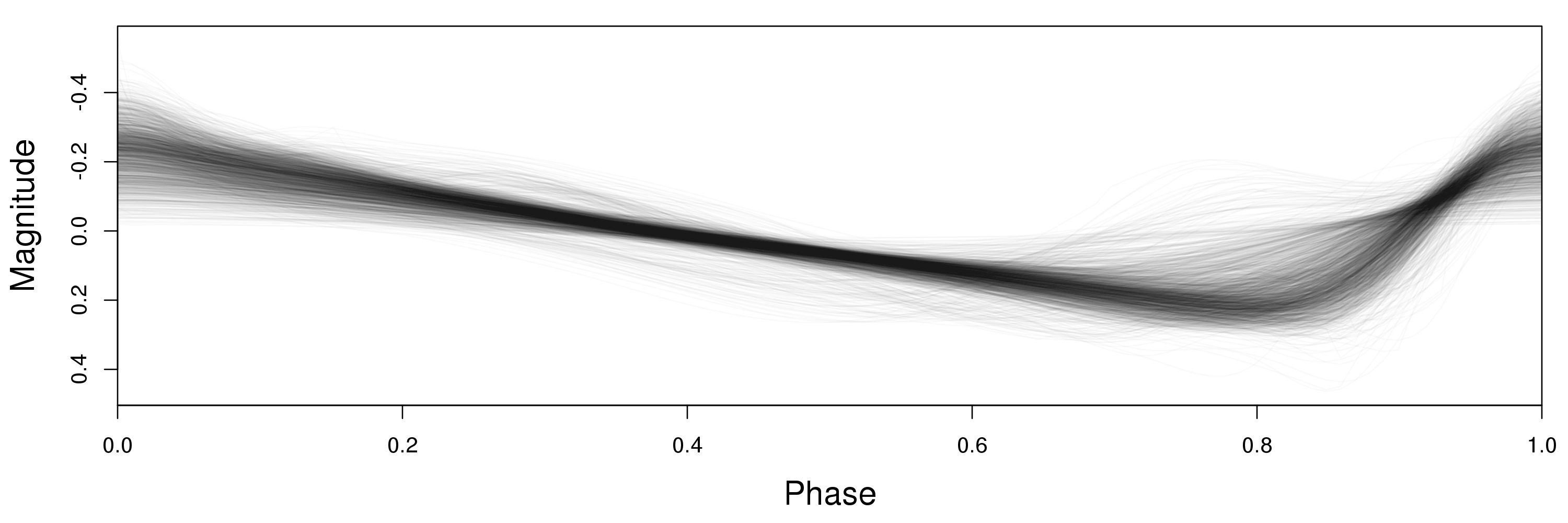}
      \caption{$1810$ folded, phase--aligned, magnitude normalized light curves of type Classical Cepheid star from the OGLE-III survey.\label{fig:lcs}}
    \end{includegraphics}
  \end{center}
\end{figure}

We use a sample of $1810$ light curves from stars of the class Classical Cepheid from the OGLE-III survey \citep{soszynski2008optical}. We normalized each light curve to mean $0$.  We smooth the light curves of each star using natural cubic splines with 15 equally spaced knots, including boundary knots at phases 0 and 1. After smoothing, we phase the smoothed light curves by assigning the minimum magnitude phase $0$. We plot the smoothed, magnitude normalized, phased light curves in Figure \ref{fig:lcs}.

\subsection{The Deepest and Least Deep Curves}

\begin{figure}[H]
  \begin{center}
    \begin{includegraphics}[scale=.4]{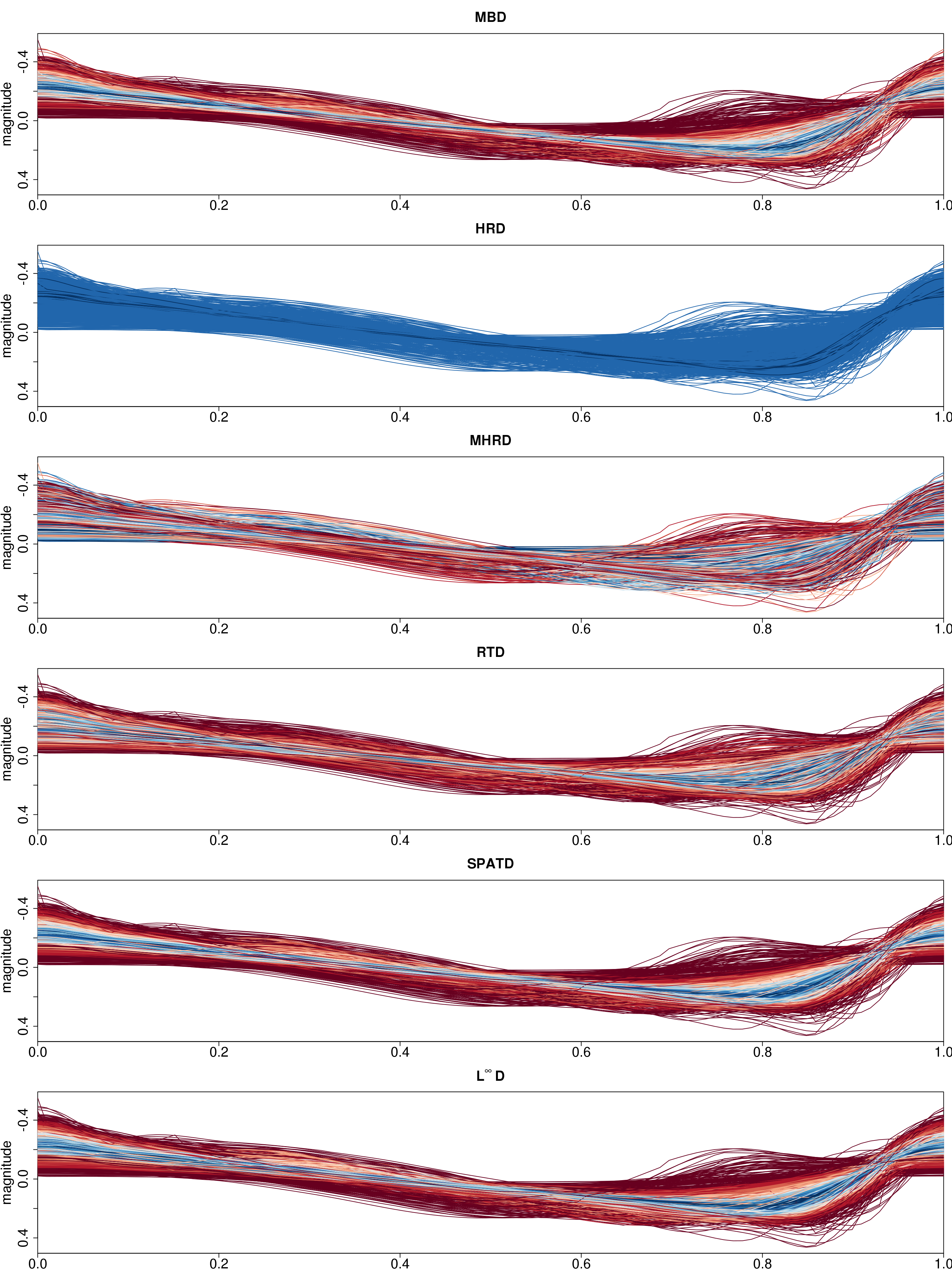}
      \caption{Functions colored by depth. Blue is deepest, red is least deep. HRD only produces 4 unique depth rankings and 93\% of functions have the minimum possible depth. The depths assigned by MHRD do not appear to represent our intuitive notions of what outliers and non--outliers are. MBD, SPATD, and $L^{\infty}D$ all produce reasonable results.\label{fig:colored}}
    \end{includegraphics}
  \end{center}
\end{figure}

Astronomers are interested in identifying the most typical and most unusual light curve shapes. Depth functions provide a method for determining typical and unusual shapes. We rank observations using HRD, MHRD, MBD, RTD, SPATD, and $L^{\infty}D$ empirical depth functions. We do not use BD because it is computationally intractable with $1810$ observations. In Figure \ref{fig:colored} we plot the lightcurves colored by the depth decile. The deepest curves are the bluest and the least deep curves are the reddest. MBD, SPATD, and $L^{\infty}D$ produce similar looking results which correspond to our intuitive notions of outliers and non--outliers. MHRD and to a lesser extent RTD produce some surprising rankings which do not correspond to intuition.

HRD only assigns 4 unique depth values to all observations. In particular, 1690 of the 1810 functions are assigned the minimum possible empirical depth of $1/1810$. Thus the ninth decile is $1/1810$ and all observations are in the top decile, resulting in all functions being colored the deepest shade of blue. This phenomenon appears to be related to issues discovered by \cite{kuelbs2014half} and \cite{chakraborty2014data} who show that for certain distributions, population HRD can be $0$ for all functions and empirical HRD may be $1/n$ for all observations. In particular, this phenomenon occurs when curves frequently cross each other, as occurs in this data set. HRD will not be useful for identifying outliers and medians in this data set. 

\begin{figure}[H]
  \begin{center}
    \begin{includegraphics}[scale=0.4]{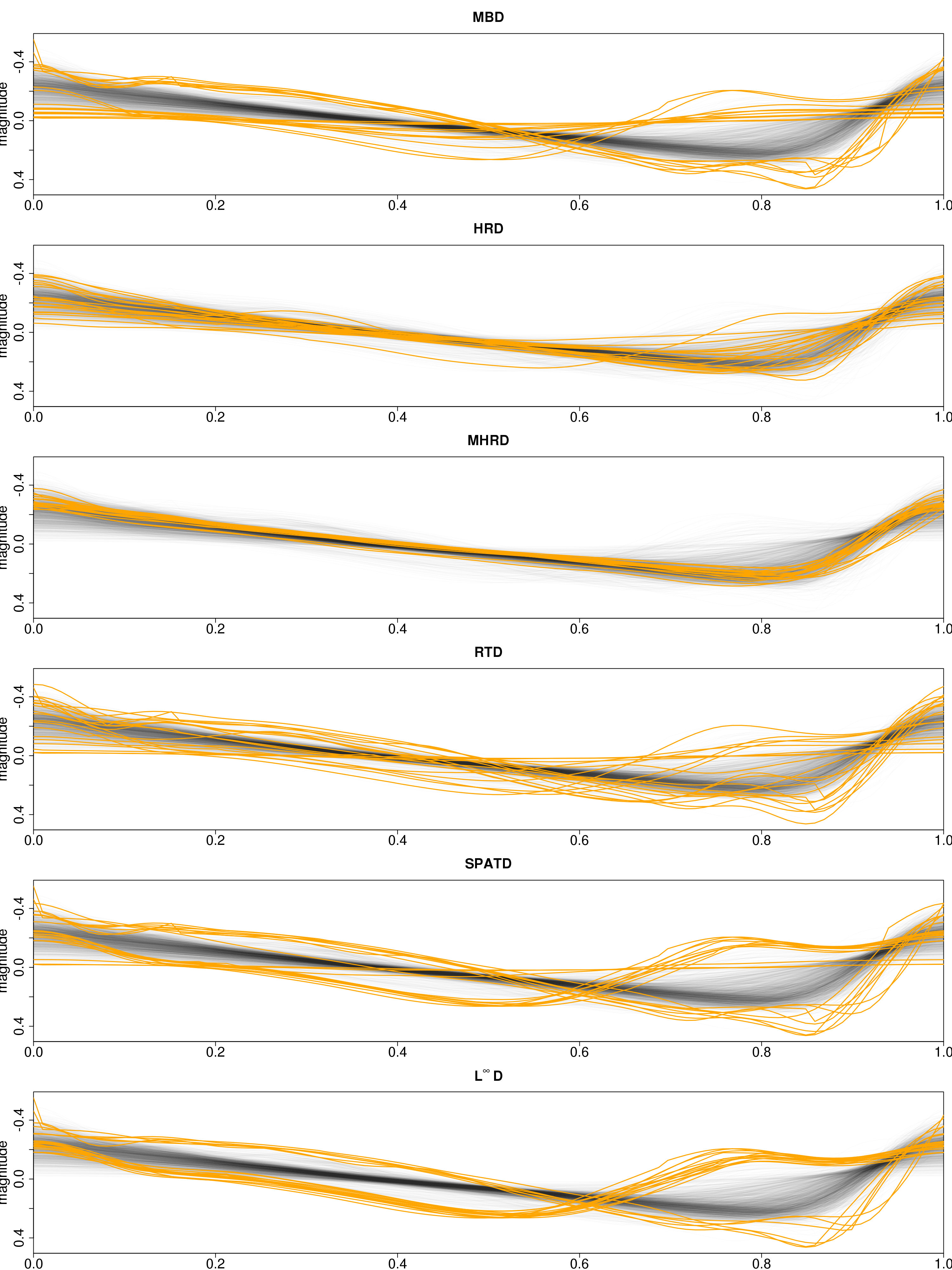}
      \caption{The 20 least deep light curves for different depth functions.\label{fig:least_deep}}
    \end{includegraphics}
  \end{center}
\end{figure}

To investigate the outliers further, in Figure \ref{fig:least_deep} we plot the 20 least deep light curves using each of the six empirical depth functions with rank ties broken at random. HRD and MHRD select light curves which do not appear to be outliers. HRD's poor performance is due to assigning too many observations the minimal possible depth. The poor behavior of MHRD is harder to explain.

\Revision{MBD, RTD, SPATD, and $L^{\infty}D$ select functions which appear to be outliers when compared with the typical shapes observed in Figure \ref{fig:lcs}. There are three distinct sets of outliers. One set of outliers have larger amplitude than other curves, with larger than average y--values values near phase 0 and smaller than average y--values at their minima near phase 0.85. A second set of outliers have small amplitude. Two of these functions are visible as SPATD outliers. A third set of outliers have a local maximum near $0.75$ where most other functions have a local minimum. These appear to be mis--registered functions. By shifting these functions in phase by $0.25$ they will align better with the other functions. Here we see an example where depth was able to detect a problem with data preprocessing.

Outliers selected by RTD appear somewhat less outlying than those selected by SPATD, MBD, and $L^\infty D$. In general SPATD produces the best set of outliers because it shows examples from all three of these sets. We note that the two low amplitude outliers detected by SPATD (and not detected by $L\infty D$) are an example of very smooth outliers among a larger set of less smooth functions. These low amplitude functions are never far in $L^\infty$ distance from the bulk of the other functions. Therefore they are not detected by $L^\infty D$. Combined with the results from the shape outlier simulation, this suggests that $L^\infty D$ is better at detecting non--smooth shape outliers in a large set of smooth functions than at detecting smooth shape outliers in a larger set of less--smooth functions.}

\Revision{Finally we note that various works have suggested incorporating measures other than simply the depth ranks to determine outlyingness. These include functional boxplots and distance measures derived from depths \citep{genton-sun-functional-box,hubert2015multivariate}. In experiments with functional boxplots, we found that they had certain advantages, such as automatically determining the number of functions to identify as outliers, and certain disadvantages, such as not detecting the low amplitude outliers.}

\subsection{Convergence of Depth Rankings}
\label{sec:depth_rankings_conv}

For many depth functions there exist limit theorems showing convergence of the empirical depth function to the population depth function. Such results were shown for $L^{\infty}D$ in Section \ref{sec:conv}. We see two limitations to these results:
\begin{enumerate}
\item In most depth applications, the observations $x_1, \ldots, x_n$ are ranked $r_1,\ldots,r_n$ where $D(x_{i},P_n) \leq D(x_{j},P_n)$ when $r_i < r_j$. Inference is then performed based on these rankings (eg robust mean estimation in Section \ref{sec:sim}). Limit theorems which show $D(\cdot,P_n) \rightarrow D(\cdot,P)$ do not directly address how empirical depth function rankings compare to population depth function rankings ie how $r_1,\ldots,r_n$ compare to $r'_1,\ldots,r'_n$ where $D(x_i,P) \leq D(x_j,P)$ when $r'_i < r'_j$.
\item Limit theorems do not indicate at what sample sizes the asymptotics take hold.
\end{enumerate}

Practitioners use the empirical depth function ranks because the population depth function is unknown. In principle one could determine the error in this approximation by plotting the rankings of the empirical depth function against the rankings of the population depth function. In practice this is impossible since the population depth function is unknown. We propose approximating this comparison by constructing two independent empirical rankings. The procedure is:
\begin{enumerate}
\item Split the data into two sets: $\vec{x}_1 = x_{1},\ldots,x_{\floor{n/2}}$ and $\vec{x}_2 = x_{\floor{n/2}+1},\ldots,x_{n}$.
\item Let $P_{n}^k$ be the empirical measure for $\vec{x}_k$ ($k=1,2)$. Construct the empirical depth functions $\mathcal{D}(\cdot,P_{n}^k)$.
\item Rank observations $\vec{x}_1$ using $\mathcal{D}(\cdot,P_{n}^1)$ to produce ranks $r_1,\ldots,r_{\floor{n/2}}$. Again rank observations $\vec{x}_1$ now using $\mathcal{D}(\cdot,P_{n}^2)$ to produce ranks $r_1',\ldots,r_{\floor{n/2}}'$.
\item Make a scatterplot of the ranking pairs $\{(r_i,r_i')\}_{i=1}^{\floor{n/2}}$.
\end{enumerate}

\begin{figure}[H]
  \begin{center}
    \begin{includegraphics}[scale=.4]{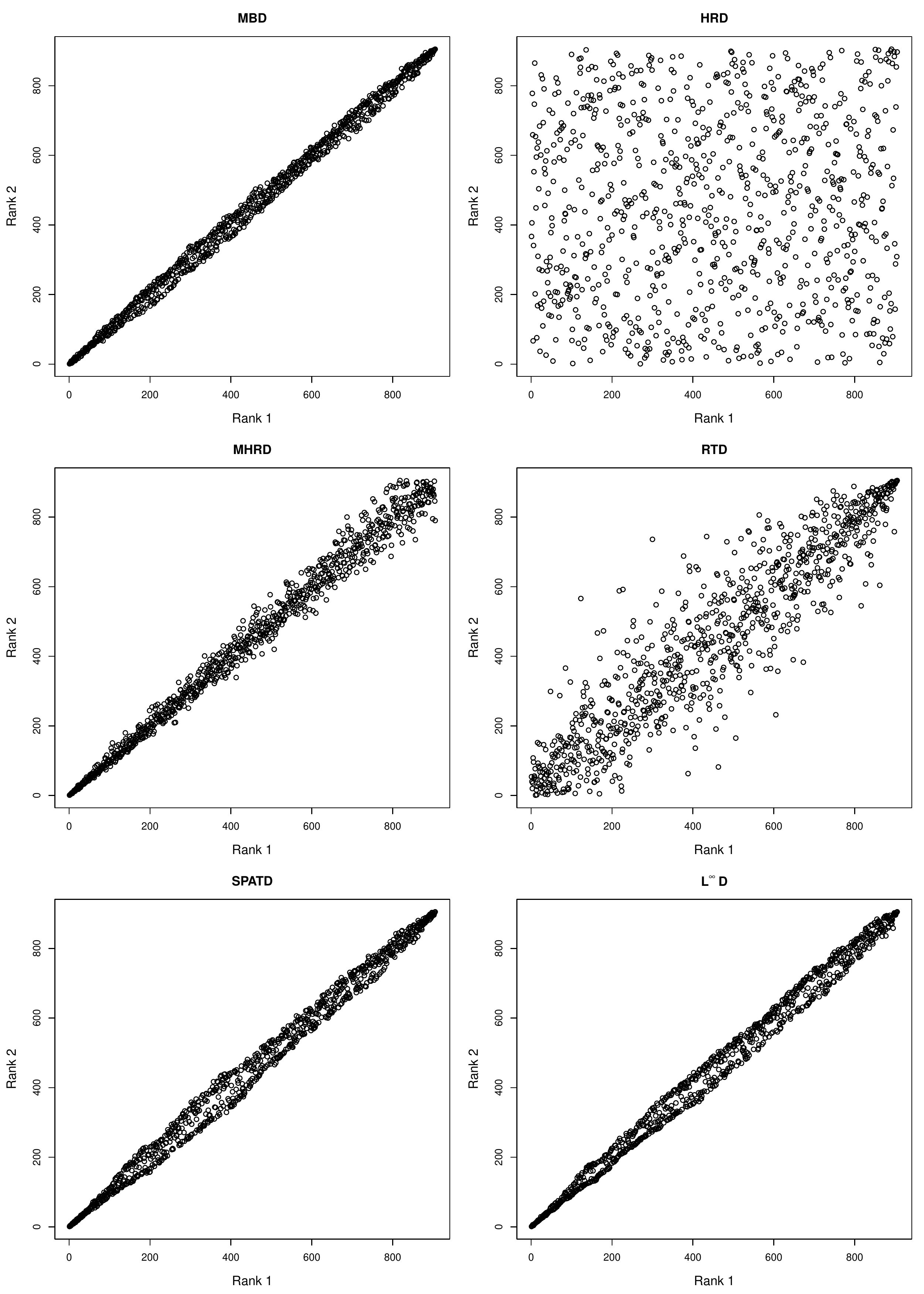}
      \caption{Rank-rank plots for six depth functions. HRD and RTD show considerable instability in rankings. There is also instability in how MHRD ranks the deepest functions. MBD, SPATD, and $L^{\infty}D$ show good stability.\label{fig:rank_scatter}}
    \end{includegraphics}
  \end{center}
\end{figure}

We term the resulting plot a rank--rank plot. Stability in these ranking can be determined by how closely the rank--rank plot follows the identity line. We perform this procedure for the light curve data using the six depth functions. Rank-rank plots are shown in Figure \ref{fig:rank_scatter}. Rank ties are broken at random. One can see immediately that HRD ranks show little agreement. This is caused by the zero depth problem. Both $\mathcal{D}(\cdot,P_{n}^1)$ and $\mathcal{D}(\cdot,P_{n}^2)$ HRD depths are zero for most functions. Since rank ties are broken at random, there is little correlation between rankings. Note that this conclusion is missed by the standard asymptotic analysis that shows $\mathcal{D}(\cdot,P_n) \rightarrow \mathcal{D}(\cdot,P)$. \Revision{RTD also shows considerable deviation from the identity line. By increasing the number of random projections we can increase the strength of the correlation. However even at 10,000 random projections there is still considerably more dispersion in ranks than with other methods (Spearman rank correlation of 0.961 for RTD with 10,000 random projections versus 0.996 for $L^\infty D$).}

The other four depth functions show much stronger correlations. For MHRD, there is some instability for the deepest ranked functions. MHRD has a maximum possible depth of $1/2$. \cite{kuelbs2014half} found that MHRD can suffer from having too many deepest functions i.e., too many functions have depth equal to $1/2$ (this is effectively the opposite of the problem experienced by HRD). It may be the case here that many of the observations have MHRD nearly equal to $1/2$, causing instability in the rankings at high depths.

To see how this result could be used, consider the practice of representing the most typical curves by selecting the deepest 1\% using an empirical depth function. How would this selection change if the population depth function were used? Using each depth function, we select the deepest 9 ($\approx 1\%$) curves as measured by $\mathcal{D}(\cdot,P_{n}^1)$. We then determine the ranking of these functions using $\mathcal{D}(\cdot,P_{n}^2)$. In Table \ref{tab:deep_ranks} we plot the $\mathcal{D}(\cdot,P_{n}^2)$ normalized rank (rank divided by sample size) of the deepest curves. Ideally these 1\% deepest curves will all have a normalized rank near 1. For MBD, SPATD, and $L^{\infty}D$, the normalized rankings are in the top 2\% of all functions. This indicates good stability of the empirical depth function and suggests that the empirical depth function ranking is accurately estimating the population depth function. For RTD, HRD, and MHRD, some functions are ranked below 90\%, suggesting instability in the rankings. 

\begin{table}[t]
\begin{center}
\input{figs/deep_ranks} 
\caption[Normalized rankings of the deepest functions]{Normalized rankings of the deepest functions.}
\label{tab:deep_ranks}
\end{center}
\end{table}

\section{Conclusions}
\label{sec:conclude}

We have shown that a simple generalization of a multivariate depth to functional data outperforms several existing notions of depth designed for functional data. \Revision{Among the depths considered, SPATD, MBD, and $L^\infty D$ appear to be the most useful. While BD, MBD, HRD, and MHRD were designed for functional data, we do not find evidence that they outperform straightforward extensions of the spatial and $L^\infty$ multivariate depths to the functional setting.}

We have discussed and critiqued the methods by which depths are evaluated. Limit theorems which show the convergence of $\mathcal{D}(\cdot,P_n)$ to $\mathcal{D}(\cdot,P)$ may not be informative as to how depth rankings behave at finite sample sizes. This is most apparent for depths such as HRD where with certain distributions the empirical depth function is converging to a population depth function which is identically $0$. The rank--rank plot of Section \ref{sec:depth_rankings_conv} may be more appropriate for determining the stability of the rankings of the empirical depth function.

\section{Supplementary Materials}
\label{sec:supp}

\begin{description}
\item[\texttt{R} code for reproducing results:] \texttt{R}--code and data needed for reproducing results in this work are available online in the archive \texttt{depth.zip}. This includes implementations of all depth functions studied and the astronomy variable star data set. The \texttt{readme} file in \texttt{depth.zip} explains how to use the code.
\end{description}

\baselineskip=14pt
\section*{Acknowledgments}

J.P.L. gratefully acknowledges support from a Texas A\&M faculty startup grant. J.Z.H. acknowledges support from NSF grant DMS-1208952. The authors are also grateful to Dr. Joel Zinn for introducing them to this research area and for helpful discussions.

\baselineskip=14pt

\bibliographystyle{abbrvnat}
\bibliography{refs}

\clearpage\pagebreak\newpage
\pagestyle{fancy}
\fancyhf{}
\rhead{\bfseries\thepage}
\lhead{\bfseries Appendix}
\begin{center}
{\LARGE{\bf Appendix to\\ {\it $L^{\infty}$ Depth for Functional Data}}}
\end{center}

\setcounter{equation}{0}
\setcounter{page}{1}
\setcounter{table}{1}
\setcounter{section}{0}
\renewcommand{\theequation}{A.\arabic{equation}}
\renewcommand{\thesection}{A.\arabic{section}}
\renewcommand{\thesubsection}{A.\arabic{section}.\arabic{subsection}}
\renewcommand{\thepage}{A.\arabic{page}}
\renewcommand{\thetable}{A.\arabic{table}}
\baselineskip=17pt

\section{Technical Notes}
\label{sec:technical_notes}

\subsection{Proof of Theorem \ref{thm:properties}}
\label{prf:properties}
\begin{itemize}[label={}]
\item P2) It is sufficient to show
\begin{equation}
\label{eq:suff}
\mathbb{E}[||\theta - X||_{\infty}] = \inf_{x \in C(I)}\mathbb{E}[||x-X||_{\infty}].
\end{equation}
Since $f(\cdot) = \mathbb{E}[||\cdot-X||_{\infty}]$ is a convex function, for all $x$ we have
\begin{equation}
\label{eq:last}
\mathbb{E}[||\theta - X||_{\infty}] \leq \frac{1}{2}\mathbb{E}[||- x + \theta - X||_{\infty}] + \frac{1}{2}\mathbb{E}[||x + \theta - X||_{\infty}]. 
\end{equation}
By assumption $\theta - X = X - \theta$ in distribution, so the RHS of \eqref{eq:last} is
\begin{align*}
\frac{1}{2}\mathbb{E}[||- x + \theta - X||_{\infty}] + \frac{1}{2}\mathbb{E}[||x - \theta +X||_{\infty}] &=\frac{1}{2}\mathbb{E}[||- x + \theta - X||_{\infty}] + \frac{1}{2}\mathbb{E}[||-x + \theta - X||_{\infty}]\\
&=\mathbb{E}[||- x + \theta - X||_{\infty}].
\end{align*}
Since the inequality holds for all $x$, we have established \eqref{eq:suff}.
\item P3) Sufficient to show
\begin{equation*}
\E[||\theta + \alpha(x-\theta) - X||_{\infty}] \leq \E[||x-X||_{\infty}]
\end{equation*}
Since $f(\cdot) = \mathbb{E}[||\cdot-X||_{\infty}]$ is a convex function we have
\begin{align*}
\E[||\theta + \alpha(x-\theta) - X||_{\infty}] &= \E[||(1-\alpha)(\theta - X) + \alpha(x - X)||_{\infty}]\\
&\leq (1-\alpha)\E[||\theta - X||_{\infty}] + \alpha\E[||x - X||_{\infty}]\\
&\leq (1-\alpha)\E[||x - X||_{\infty}] + \alpha\E[||x - X||_{\infty}]\\
&= \E[||x - X||_{\infty}]
\end{align*}
\item P4) Sufficient to show that $\mathbb{E}[||x-X||_{\infty}] \rightarrow \infty$ as $||x||_{\infty} \rightarrow \infty$. To see this, note that by the triangle inequality $||x||_{\infty} - \mathbb{E}[||X||_{\infty}] \leq \mathbb{E}[||x-X||_{\infty}]$.
\end{itemize}
If $A$ is an isometry with respect to the $L^{\infty}$ norm ($||A(x)||_{\infty} = ||x||_\infty \, \, \forall x$) then it is a linear map so
\begin{equation*}
L^\infty D(x,P) = \frac{1}{1+\E[||x-X||_\infty]} = \frac{1}{1+\E[||A(x)-A(X)||_\infty]} = L^\infty D(A(x),P_{A(X)}).
\end{equation*}

\subsection{Proof of Theorem \ref{thm:slln}}
\label{prf:slln}
Let $f(\cdot) = (1 + \cdot)^{-1}$. By the Strong Law of Large Numbers $n^{-1} \sum_{i=1}^n ||X_i - x||_{\infty} \rightarrow \mu_x \geq 0$. By the continuous mapping theorem
\begin{equation*}
L^{\infty}(x,P_n) = f(\frac{1}{n}\sum||X_i - x||_{\infty}) \rightarrow_{a.s.} f(\mu_x) = L^{\infty}(x,P).
\end{equation*}

\subsection{Proof of Theorem \ref{thm:clt}}
\label{prf:clt}
Let $f(\cdot) = (1 + \cdot)^{-1}$. By the CLT $\sqrt{n}(n^{-1}\sum ||X_i-x||_{\infty} - \mu_x) \rightarrow N(0,\sigma_x^2)$. Apply the delta method with $f$ to obtain
\begin{align*}
\sqrt{n}(L^{\infty}(x,P_n) - L^{\infty}(x,P)) &= \sqrt{n}(f(n^{-1}\sum ||X_i-x||_{\infty}) - f(\mu_x))\\
&\rightarrow N(0,f'(\mu_x)^2\sigma_x^2)\\
&= N(0,\sigma_x^2(1+\mu_x)^{-4}).
\end{align*}

\section{Computational Speed}
\label{sec:computational_speed}

\begin{figure}[h]
  \begin{center}
    \begin{includegraphics}[scale=0.5]{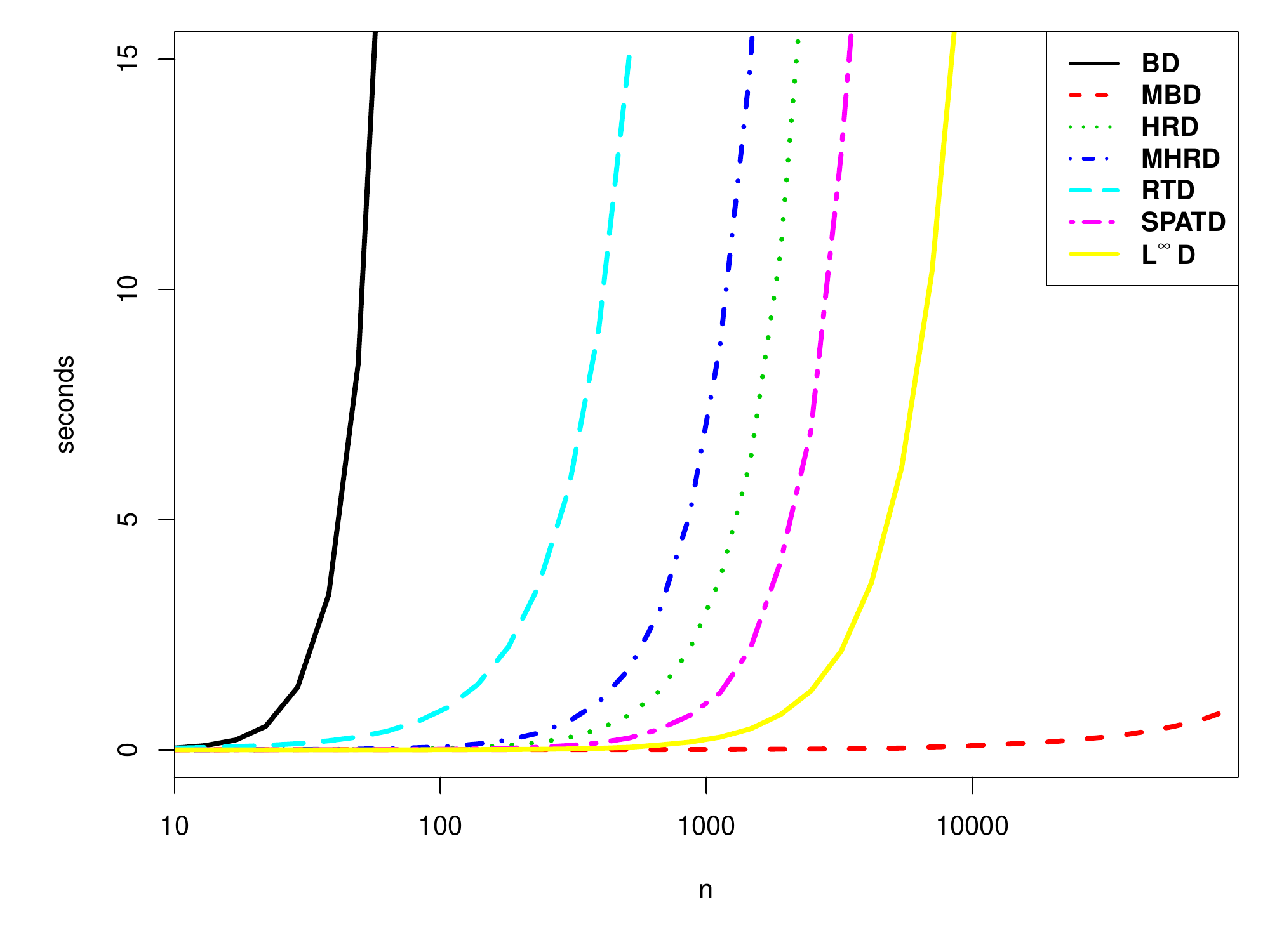}
      \caption{Comparison of computation time for several depth functions.\label{figs:depths_speed}}
    \end{includegraphics}
  \end{center}
\end{figure}

We determine the time to compute all $n$ sample observation depths for each depth function. All functions are sampled at $50$ time points. Code is written in \texttt{R} and run on an Intel(R) Core(TM) i7-3770 CPU at 3.40GHz. BD (black solid) is the slowest and becomes computationally infeasible at $n$ less than 100. This slow computation time is due to the fact that BD with $J=3$ is an $O(n^4)$ algorithm \citep{kwon2016clustering}. HRD, MHRD, Spatial, and $L^{\infty}D$ are all $O(n^2)$ algorithms with similar computation times. The speed improvement of $L^{\infty}D$ over the other methods is due to our use of distance matrices in the \texttt{R} code. RTD is $O(n^2N_v)$ where $N_v$ is the number of random projections. While typically $N_v$ is fixed with $n$, \cite{cuesta2008random} recommend a value of $250$ which makes the algorithm significantly slower than other $O(n^2)$ depths. MBD with $J=2$ is extremely fast due to an algorithm of \cite{sun2012exact} which allows all sample function depths to be computed in $O(n\log n)$ time \citep{kwon2016clustering}.

\end{document}

%% file: figs/mag_M5_table.tex
\begin{table}[ht]
\centering
\begin{tabular}{rlllll}
  \hline
 & M0 & M1 & M2 & M3 & M4 \\ 
  \hline
MEAN & \textbf{0.0196 (0.0009)} & 0.3054 (0.0114) & 0.0745 (0.0047) & 0.0472 (0.0023) & \textbf{0.0258 (0.0010)} \\ 
  MED & 0.0301 (0.0011) & 0.0581 (0.0025) & 0.0383 (0.0017) & 0.0349 (0.0014) & 0.0331 (0.0012) \\ 
  BD & 0.0268 (0.0013) & 0.2803 (0.0117) & 0.0702 (0.0052) & 0.0442 (0.0023) & 0.0290 (0.0013) \\ 
  MBD & 0.0258 (0.0012) & 0.0611 (0.0036) & 0.0319 (0.0017) & 0.0365 (0.0017) & 0.0330 (0.0013) \\ 
  HRD & 0.0298 (0.0014) & 0.3208 (0.0139) & 0.0762 (0.0052) & 0.0580 (0.0029) & 0.0331 (0.0015) \\ 
  MHRD & 0.0266 (0.0013) & \textbf{0.0548 (0.0032)} & \textbf{0.0315 (0.0017)} & 0.0500 (0.0024) & 0.0336 (0.0013) \\ 
  RTD & 0.0252 (0.0012) & 0.2979 (0.0123) & 0.0853 (0.0056) & 0.0585 (0.0034) & 0.0323 (0.0014) \\ 
  SPATD & \textbf{0.0241 (0.0011)} & 0.0748 (0.0044) & 0.0327 (0.0017) & \textbf{0.0297 (0.0014)} & \textbf{0.0270 (0.0011)} \\ 
  $L^{\INFTY}D$ & 0.0287 (0.0014) & \textbf{0.0328 (0.0017)} & \textbf{0.0305 (0.0015)} & \textbf{0.0316 (0.0015)} & 0.0316 (0.0015) \\ 
   \hline
\end{tabular}
\caption{MISE for mean estimation with M=5 level of magnitude contamination. The probability a curve is contaminated is q=0.1.} 
\label{figs/mag_M5}
\end{table}

%% file: figs/mag_M25_table.tex
\begin{table}[ht]
\centering
\begin{tabular}{rlllll}
  \hline
 & M0 & M1 & M2 & M3 & M4 \\ 
  \hline
MEAN & \textbf{0.0182 (0.0008)} & 7.1273 (0.2409) & 1.1960 (0.0734) & 0.6688 (0.0361) & 0.1504 (0.0039) \\ 
  MED & 0.0299 (0.0011) & \textbf{0.0576 (0.0026)} & \textbf{0.0368 (0.0014)} & \textbf{0.0340 (0.0013)} & \textbf{0.0310 (0.0011)} \\ 
  BD & 0.0251 (0.0012) & 3.7588 (0.2011) & 0.5689 (0.0508) & 0.2337 (0.0177) & 0.1014 (0.0039) \\ 
  MBD & 0.0235 (0.0012) & 0.3826 (0.0499) & 0.0561 (0.0097) & 0.1951 (0.0121) & 0.1819 (0.0049) \\ 
  HRD & 0.0255 (0.0012) & 4.3661 (0.2348) & 0.3789 (0.0391) & 0.5863 (0.0359) & 0.1490 (0.0046) \\ 
  MHRD & 0.0247 (0.0012) & 0.2874 (0.0398) & 0.0525 (0.0092) & 0.4205 (0.0249) & 0.1954 (0.0049) \\ 
  RTD & 0.0248 (0.0011) & 5.4750 (0.2241) & 0.6924 (0.0543) & 0.2360 (0.0197) & 0.0921 (0.0031) \\ 
  SPATD & \textbf{0.0234 (0.0011)} & 0.5729 (0.0641) & 0.0565 (0.0098) & \textbf{0.0302 (0.0021)} & \textbf{0.0275 (0.0016)} \\ 
  $L^{\INFTY}D$ & 0.0287 (0.0014) & \textbf{0.0322 (0.0015)} & \textbf{0.0438 (0.0085)} & 0.0362 (0.0059) & 0.0312 (0.0017) \\ 
   \hline
\end{tabular}
\caption{MISE for mean estimation with M=25 level of magnitude contamination. The probability a curve is contaminated is q=0.1.} 
\label{figs/mag_M25}
\end{table}

%% file: figs/shape_table.tex
\begin{table}[ht]
\centering
\begin{tabular}{rlllll}
  \hline
 & M5 & M6 & M7 & M8 & M9 \\ 
  \hline
MEAN & \textbf{0.0224 (0.0010)} & \textbf{0.0187 (0.0010)} & \textbf{0.0203 (0.0011)} & \textbf{0.0205 (0.0012)} & \textbf{0.0200 (0.0011)} \\ 
  MED & 0.0339 (0.0013) & 0.0297 (0.0013) & 0.0311 (0.0015) & 0.0310 (0.0014) & 0.0282 (0.0012) \\ 
  BD & 0.0280 (0.0014) & 0.0250 (0.0014) & 0.0276 (0.0016) & 0.0265 (0.0014) & 0.0266 (0.0016) \\ 
  MBD & 0.0272 (0.0012) & \textbf{0.0223 (0.0011)} & \textbf{0.0244 (0.0013)} & 0.0240 (0.0012) & 0.0232 (0.0013) \\ 
  HRD & 0.0312 (0.0016) & 0.0265 (0.0014) & 0.0286 (0.0016) & 0.0288 (0.0016) & 0.0283 (0.0017) \\ 
  MHRD & 0.0274 (0.0012) & 0.0226 (0.0011) & 0.0249 (0.0013) & \textbf{0.0237 (0.0012)} & 0.0234 (0.0012) \\ 
  RTD & 0.0287 (0.0014) & 0.0246 (0.0013) & 0.0272 (0.0015) & 0.0269 (0.0015) & 0.0264 (0.0014) \\ 
  SPATD & \textbf{0.0264 (0.0012)} & 0.0228 (0.0012) & 0.0246 (0.0014) & 0.0248 (0.0013) & \textbf{0.0231 (0.0012)} \\ 
  $L^{\INFTY}D$ & 0.0323 (0.0016) & 0.0286 (0.0017) & 0.0289 (0.0017) & 0.0286 (0.0016) & 0.0273 (0.0014) \\ 
   \hline
\end{tabular}
\caption{MISE for mean estimation in shape contamination models. The probability a curve is contaminated is q=0.15.} 
\label{figs/shape}
\end{table}

%% file: figs/detect.tex
\begin{table}[ht]
\centering
\begin{tabular}{rlllll}
  \hline
 & M5 & M6 & M7 & M8 & M9 \\ 
  \hline
BD & \textbf{0.912} & \textbf{0.956} & \textbf{0.92} & \textbf{0.952} & \textbf{0.952} \\ 
  MBD & 0.096 & 0.074 & 0.08 & 0.058 & 0.066 \\ 
  HRD & 0.434 & 0.472 & 0.45 & 0.438 & 0.46 \\ 
  MHRD & 0.07 & 0.038 & 0.058 & 0.038 & 0.044 \\ 
  RTD & 0.024 & 0.024 & 0.03 & 0.026 & 0.024 \\ 
  SPATD & 0.194 & 0.198 & 0.19 & 0.166 & 0.168 \\ 
  $L^{\INFTY}D$ & \textbf{0.98} & \textbf{0.996} & \textbf{0.988} & \textbf{0.986} & \textbf{0.986} \\ 
   \hline
\end{tabular}
\caption{The fraction of times the outlier was ranked in the least deep 0.2 fraction of functions in the shape contamination models.} 
\label{fig:detect}
\end{table}

%% file: figs/deep_ranks.tex
\begin{tabular}{ll}
  \hline
Depth & Normalized Ranks \\ 
  \hline
MBD & 0.983, 0.99, 0.991, 0.992, 0.996, 0.997, 0.998, 0.999, 1 \\ 
  HRD & 0.049, 0.128, 0.188, 0.221, 0.472, 0.901, 0.95, 0.957, 0.996 \\ 
  MHRD & 0.872, 0.878, 0.935, 0.952, 0.957, 0.959, 0.967, 0.972, 0.998 \\ 
  RTD & 0.833, 0.988, 0.99, 0.992, 0.994, 0.996, 0.998, 0.999, 1 \\ 
  SPATD & 0.988, 0.99, 0.993, 0.994, 0.996, 0.997, 0.998, 0.999, 1 \\ 
  $L^{\infty}D$ & 0.99, 0.991, 0.993, 0.994, 0.996, 0.997, 0.998, 0.999, 1 \\ 
   \hline
\end{tabular}